# Atlas-Based Segmentation of Intracochlear Anatomy in Metal Artifact Affected CT Images of the Ear with Co-trained Deep Neural Networks

Jianing Wang, Dingjie Su, Yubo Fan, Srijata Chakravorti, Jack H. Noble, and Benoit M. Dawant

Dept. of Electrical and Computer Engineering
Vanderbilt University, Nashville, TN 37235, USA

**Abstract.** We propose an atlas-based method to segment the intracochlear anatomy (ICA) in the post-implantation CT (Post-CT) images of cochlear implant (CI) recipients that preserves the point-to-point correspondence between the meshes in the atlas and the segmented volumes. To solve this problem, which is challenging because of the strong artifacts produced by the implant, we use a pair of co-trained deep networks that generate dense deformation fields (DDFs) in opposite directions. One network is tasked with registering an atlas image to the Post-CT images and the other network is tasked with registering the Post-CT images to the atlas image. The networks are trained using loss functions based on voxel-wise labels, image content, fiducial registration error, and cycle-consistency constraint. The segmentation of the ICA in the Post-CT images is subsequently obtained by transferring the predefined segmentation meshes of the ICA in the atlas image to the Post-CT images using the corresponding DDFs generated by the trained registration networks. Our model can learn the underlying geometric features of the ICA even though they are obscured by the metal artifacts. We show that our end-to-end network produces results that are comparable to the current state of the art (SOTA) that relies on a two-steps approach that first uses conditional generative adversarial networks to synthesize artifact-free images from the Post-CT images and then uses an active shape model-based method to segment the ICA in the synthetic images. Our method requires a fraction of the time needed by the SOTA, which is important for end-user acceptance.

## 1    Introduction

The cochlea (Figure 1c) is a spiral-shaped structure that is part of the inner ear involved in hearing. It contains two main cavities: the scala tympani (ST) and the scala vestibuli (SV). The modiolus (MD) is a porous bone around which the cochlea is wrapped that hosts the auditory nerves. A cochlear implant (CI) is an implanted neuroprosthetic device that is designed to produce hearing sensations in a person with severe to profound deafness by electrically stimulating the auditory nerves [1]. CIs are programmed postoperatively in a process that involves activating all or a subset of the electrodes and adjusting the stimulus level for each of these to a level that is beneficial to the recipient [2]. Programming parameters adjustment is influenced by the intracochlear position of the CI electrodes, which requires the accurate localization of the CI electrodes relative to the intracochlear anatomy (ICA) in the post-implantation CT (Post-CT) images of the CI recipients. This, in turn, requires the accurate segmentation of the ICA in the Post-CT images. Segmenting the ICA in the Post-CT images is challenging due to the strong artifacts produced by the metallic CI electrodes (Figure 1b) that can obscure these structures, often severely. For patients who have been scanned before implantation, the segmentation of the ICA can be obtained by segmenting their pre-implantation CT (Pre-CT) image (Figure 1a) using an active shape model-based (ASM) method [3]. The outputs of the ASM method are surface meshes of the ST, the SV, and the MD that have a predefined number of vertices. Importantly, each vertex corresponds to a specific anatomical location on the surface of the structures and the meshes are encoded with the information needed for the programming of the implant. Preserving point-to-point correspondence when registering the images is thus of critical importance in our application. The ICA in the Post-CT image of the patients can be obtained by registering their Pre-CT image to the Post-CT image and then transferring the segmentations of the ICA in the Pre-CT image to the Post-CT image using that transformation. This approach does not extend to CI recipients for whom a Pre-CT image is unavailable, which is the case for long-term recipients who were not scanned before surgery, or for recipients for whom images cannot be retrieved. To overcome this issue, Wang *et al.* have proposed a two-step method [4, 5], which we refer to as "cGANs+ASM". The method first uses conditional generative adversarial networks (cGANs) [6, 7] to synthesize artifact-free Pre-CT images from the Post-CT images and then uses the ASM method [3] to segment the ICA in the synthetic images. To the best of our knowledge, cGANs+ASM is the most accurate published automatic method for ICA segmentation in Post-CT images.

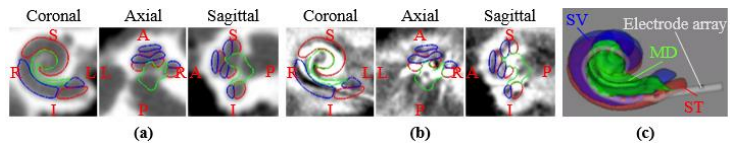

**Fig. 1.** A pair of registered (a) Pre-CT and (b) Post-CT images of an ear of a CI recipient. (c) An illustration of the intracochlear anatomy with an implanted CI electrode array. The meshes of the ST, the SV, and the MD are obtained by applying the ASM method to the Pre-CT image.

Here, we propose an end-to-end atlas-based method: we first generate a dense deformation field (DDF) between an artifact-free atlas image and a Post-CT image. The segmentation of the ICA in the Post-CT image can then be obtained by transferring the predefined segmentation meshes of the ICA in the atlas image to the Post-CT image using that DDF. We note that the inter-subject non-rigid registration between the atlas image and the Post-CT image is a difficult task because (1) considerable variation in cochlear anatomy across individuals has been documented [8], and (2) the artifacts in the Post-CT image change, often severely, the appearance of the anatomy, which has a significant influence on the accuracy of registration methods guided by intensity-based similarity metrics. To overcome the challenges, we propose a method to perform registrations between an atlas image and the Post-CT images that rely on deep networks. Following the idea of consistent image registration obtained by jointly estimating the forward and reverse transformations between two images that is proposed by Christensen *et al.* [9], we use a pair of co-trained networks that generate DDFs in opposite directions. One network is tasked with registering the atlas image to the Post-CT image and the other one is tasked with registering the Post-CT image to the atlas image. The networks are trained using loss functions that include voxel-wise labels, image content, fiducial registration error (FRE), and cycle-consistency constraint. We show that our model can segment the ICA and preserve point-to-point correspondence between the atlas and the Post-CT meshes, even when the ICA is difficult to localize visually.

## 2 Method

### 2.1 Data

Our dataset consists of Pre-CT and Post-CT image pairs of 624 ears. The atlas image is a Pre-CT image of an ear that is not in the 624 ears. The Pre-CT images are acquired with several conventional scanners (GE BrightSpeed, LightSpeed Ultra; Siemens Sensation 16; and Philips Mx8000 IDT, iCT 128, and Brilliance 64) and the Post-CT images are acquired with a low-dose flat-panel volumetric scanner (Xoran Technologies xCAT® ENT). The typical voxel size is 0.25×0.25×0.3mm$^3$ for the Pre-CT images and 0.4×0.4×0.4mm$^3$ for the Post-CT images. For each ear, the Pre-CT image is rigidly registered to the Post-CT image. The registration is accurate because the surgery, which consists of threading an electrode array through a small hole into the bony cavity, does not induce non-rigid deformation of the cochlea. The registered Pre-CT and Post-CT image pairs are then aligned to the atlas image so that the ears are roughly in the same spatial location and orientation. All of the images are resampled to an isotropic voxel size of 0.2mm. Images of 64×64×64 voxels that contain the cochleae are cropped from the full-sized images, and our networks are trained to process such cropped images.

### 2.2 Learning to Register the Artifact-affected Images and the Atlas Image with Assistance of the Paired Artifact-free Images

Figure 2a shows a list of images, meshes, and masks used to train our networks. For simplicity, we use $O_{xSpc}$ to denote an object $O$ in the $x$ space. For example, $AtlasImg_{atlasSpc}$ is our atlas image in the atlas space. Similarly, $PostImg_{postSpc}$ is a Post-CT image in the Post-CT space. $Mesh_{atlasSpc}$ is the segmentation mesh of the ICA in $AtlasImg_{atlasSpc}$ generated by applying the ASM method to $AtlasImg_{atlasSpc}$. $PreImg_{postSpc}$ is the paired Pre-CT image of $PostImg_{postSpc}$ registered to the original Post-CT. $Mesh_{postSpc}$ is the segmentation mesh of the ICA in $PostImg_{postSpc}$. It has been generated by applying the ASM method to $PreImg_{postSpc}$ and then transferring the meshes to $PostImg_{postSpc}$. $Mask_{atlasSpc}$ and $Mask_{postSpc}$ are segmentation masks of the ST, SV, and MD. They are generated by converting $Mesh_{atlasSpc}$ and $Mesh_{postSpc}$ to masks.

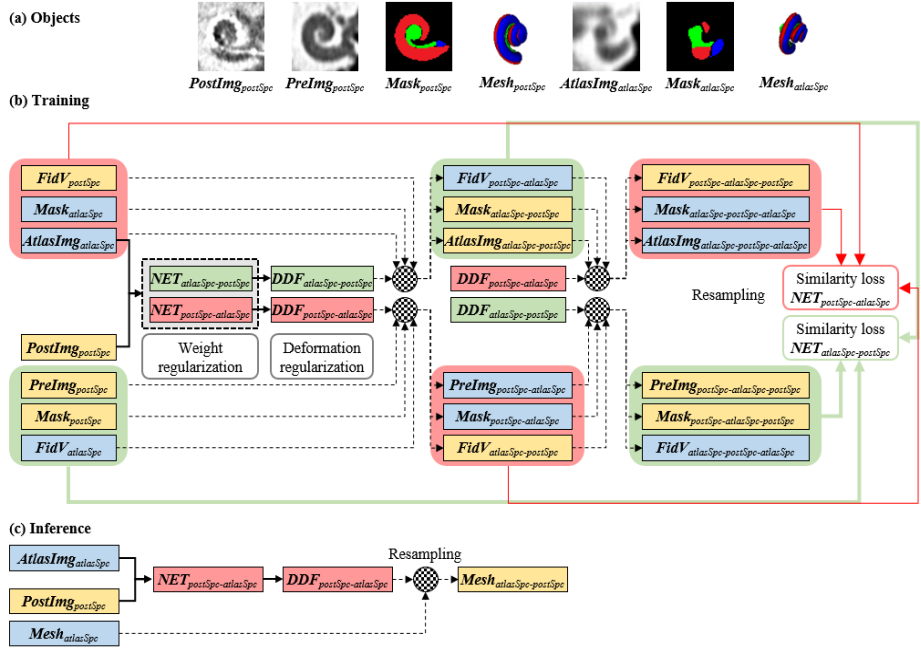

**Fig. 2.** The framework of our method. (a) Objects used for training the networks. (b) Training phase. (c) Inference phase.

As shown in Figure 2b, the input of our networks is the concatenation of **AtlasImg**$_{atlasSpc}$ and **PostImg**$_{postSpc}$. The networks consist of a first network (**NET**$_{atlasSpc\text{-}postSpc}$) that generates a DDF from the atlas space to the Post-CT space (**DDF**$_{atlasSpc\text{-}postSpc}$) and a second network (**NET**$_{postSpc\text{-}atlasSpc}$) that generates a DDF from the Post-CT space to the atlas space (**DDF**$_{postSpc\text{-}atlasSpc}$). **FidV**$_{atlasSpc}$ and **FidV**$_{postSpc}$ are fiducial vertices randomly sampled from **Mesh**$_{atlasSpc}$ and **Mesh**$_{postSpc}$ on the fly for calculating FRE during training.

Assuming that *sSpc* is the source space and *tSpc* is the target space. The Pre-CT image, the segmentation masks, and the fiducial points in *sSpc* are warped to *tSpc* by using the corresponding DDFs (note that one DDF is used for the images and masks and the other for the fiducial points), and the results are denoted as **PreImg**$_{sSpc\text{-}tSpc}$, **Mask**$_{sSpc\text{-}tSpc}$, and **FidV**$_{sSpc\text{-}tSpc}$. Then, **PreImg**$_{sSpc\text{-}tSpc}$, **Mask**$_{sSpc\text{-}tSpc}$, and **FidV**$_{sSpc\text{-}tSpc}$ are transferred back to *sSpc* using the corresponding DDF, and the results are denoted as **PreImg**$_{sSpc\text{-}tSpc\text{-}sSpc}$, **Mask**$_{sSpc\text{-}tSpc\text{-}sSpc}$, and **FidV**$_{sSpc\text{-}tSpc\text{-}sSpc}$, respectively. The training objective for **NET**$_{sSpc\text{-}tSpc}$ can be constructed by using similarity measurements between the target object in *tSpc* (denoted as $O_{tSpc}$) and the source object that has been transferred to *tSpc* from *sSpc* (denoted as $O_{sSpc\text{-}tSpc}$). Specifically, we use the multiscale soft probabilistic Dice (MSPDice) [10] between **Mask**$_{tSpc}$ and **Mask**$_{sSpc\text{-}tSpc}$, which is denoted as MSPDice(**Mask**$_{tSpc}$, **Mask**$_{sSpc\text{-}tSpc}$), to measure the similarity of the segmentation masks. The multiscale soft probabilistic Dice is less sensitive to the class imbalance in the segmentation tasks and is more appropriate for measuring label similarity in the context of image registration [11]. The similarity between **FidV**$_{tSpc}$ and **FidV**$_{sSpc\text{-}tSpc}$ is measured by the mean fiducial registration error $\overline{\text{FRE}}(\textbf{\textit{FidV}}_{tSpc}, \textbf{\textit{FidV}}_{sSpc\text{-}tSpc})$, which is calculated as the average Euclidean distance between the vertices in $\textbf{\textit{FidV}}_{tSpc}$ and the corresponding vertices in $\textbf{\textit{FidV}}_{sSpc\text{-}tSpc}$. The Post-CT images cannot be used for calculating intensity-based loss due to the artifacts, thus we use the normalized cross-correlation (NCC) between $\textbf{\textit{PreImg}}_{tSpc}$ and $\textbf{\textit{PreImg}}_{sSpc\text{-}tSpc}$, which is denoted as NCC($\textbf{\textit{PreImg}}_{tSpc}$, $\textbf{\textit{PreImg}}_{sSpc\text{-}tSpc}$), to measure the similarity between the warped source image and the target image. A cycle-consistency loss is used for regularizing the transformations. It imposes inverse consistency between the objects in the two spaces and has been shown to reduce folding problems [12]. Our cycle-consistency loss $\textbf{\textit{CycConsis}}_{sSpc\text{-}tSpc}$ measures the similarity between the original source objects in the source space and the source objects that have been transferred from the source space to the target space and then transferred back to the source space, it is calculated as MSPDice($\textbf{\textit{Mask}}_{sSpc}$, $\textbf{\textit{Mask}}_{sSpc\text{-}tSpc\text{-}sSpc}$) + 2×$\overline{\text{FRE}}$($\textbf{\textit{FidV}}_{sSpc}$, $\textbf{\textit{FidV}}_{sSpc\text{-}tSpc\text{-}sSpc}$) + 0.5×NCC($\textbf{\textit{PreImg}}_{sSpc}$, $\textbf{\textit{PreImg}}_{sSpc\text{-}tSpc\text{-}sSpc}$). Furthermore, the DDF from the source space to the target space $\textbf{\textit{DDF}}_{sSpc\text{-}tSpc}$ is regularized using bending energy [13], which is denoted as BendE($\textbf{\textit{DDF}}_{sSpc\text{-}tSpc}$). The learnable parameters of the registration network $\textbf{\textit{NET}}_{sSpc\text{-}tSpc}$ (except for the biases) are regularized by an L2 term, which is denoted as L2($\textbf{\textit{NET}}_{sSpc\text{-}tSpc}$). To summarize, the training objective for our networks is the weighted sum of the loss terms listed in Table 1; wherein the weights have been selected empirically by looking at training performance on a small number of epochs.

**Table 1.** Loss terms that are used to train our model.

| Loss | Definition | Weight |
|---|---|---|
| MSPDice | MSPDice($\textbf{\textit{Mask}}_{postSpc}$, $\textbf{\textit{Mask}}_{atlasSpc\text{-}postSpc}$) + MSPDice($\textbf{\textit{Mask}}_{atlasSpc}$, $\textbf{\textit{Mask}}_{postSpc\text{-}atlasSpc}$) | 1 |
| Mean FRE | $\overline{\text{FRE}}(\textbf{\textit{FidV}}_{postSpc}, \textbf{\textit{FidV}}_{atlasSpc\text{-}postSpc})$ + $\overline{\text{FRE}}(\textbf{\textit{FidV}}_{atlasSpc}, \textbf{\textit{FidV}}_{postSpc\text{-}atlasSpc})$ | 2 |
| NCC | NCC($\textbf{\textit{PreImg}}_{postSpc}$, $\textbf{\textit{AtlasImg}}_{atlasSpc\text{-}postSpc}$) + NCC($\textbf{\textit{AtlasImg}}_{atlasSpc}$, $\textbf{\textit{PreImg}}_{postSpc\text{-}atlasSpc}$) | 0.5 |
| Cycle-consistency | $\textbf{\textit{CycConsis}}_{atlasSpc\text{-}postSpc}$ + $\textbf{\textit{CycConsis}}_{postSpc\text{-}atlasSpc}$ | 0.5 |
| BendE | BendE($\textbf{\textit{DDF}}_{atlasSpc\text{-}postSpc}$) + BendE($\textbf{\textit{DDF}}_{postSpc\text{-}atlasSpc}$) | 0.5 |
| L2 | L2($\textbf{\textit{NET}}_{atlasSpc\text{-}postSpc}$) + L2($\textbf{\textit{NET}}_{postSpc\text{-}atlasSpc}$) | 0.0001 |

### 2.3 Network Architecture

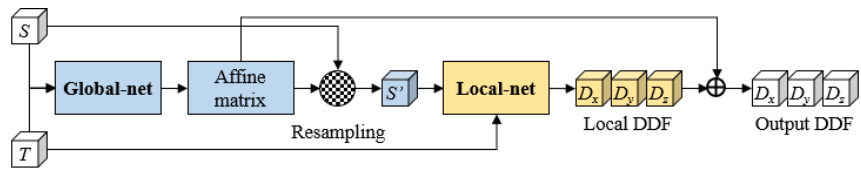

**Fig. 3.** Illustration of a registration network $\textbf{\textit{NET}}_{sSpc\text{-}tSpc}$ that is tasked to generate a DDF from the source space to the target space.

The registration networks in our model are adapted from the network architecture proposed by Hu *et al*. [14] and Ghavami *et al*. [15]. As shown in Figure 3, ***NET**$_{sSpc\text{-}tSpc}$*, which is tasked with generating a DDF for warping the source image *S* to the target image *T*, is composed of a Global-net and a Local-net. After receiving the concatenation of *S* and *T*, the Global-net generates an affine transformation matrix. *S* is warped to *T* by using this affine transformation and the resulting image is denoted as *S'*. Then, the Local-net takes the concatenation of *S'* and *T* to generate a non-rigid local DDF. The affine transformation and the local DDF are composed to produce the output DDF. The details about the Global-net and Local-net can be found in [14].

### 2.4 Evaluation

As shown in Figure 2c, at the inference phase, given a new Post-CT image ***PostImg**$_{postSpc}$*, the ICA in ***PostImg**$_{postSpc}$* can be segmented by warping ***Mesh**$_{atlasSpc}$* to ***PostImg**$_{postSpc}$* using the DDF generated by the trained network. The resulting segmentation mesh of the ICA is denoted as ***Mesh**$_{atlasSpc\text{-}postSpc}$*. ***Mesh**$_{postSpc}$*, which has been described in Section 2.2, is used as the ground truth for comparison. As ***Mesh**$_{atlasSpc}$* and ***Mesh**$_{postSpc}$* are the outputs of the ASM method, both of them have a predefined number of vertices, and the vertices of ***Mesh**$_{atlasSpc}$* and ***Mesh**$_{postSpc}$* have a one-to-one correspondence. There are 3344, 3132, and 2852 vertices on the ST, SV, and MD mesh surfaces, respectively, for a total of 9328 vertices. Point-to-point error (P2PE), computed as the Euclidean distance in millimeters, between the corresponding vertices on ***Mesh**$_{atlasSpc\text{-}postSpc}$* and ***Mesh**$_{postSpc}$* are used to quantify the accuracy of the segmentation and registration. The P2PEs between the corresponding vertices on ***Mesh**$_{postSpc}$* and the meshes generated by cGANs+ASM are calculated and serve as values that are used to compare the proposed method with the state of the art (SOTA). The method proposed in [14], which uses a unidirectional registration network trained with the MSPDice loss and the regularization loss, is used as a baseline for comparison. In addition to the MSPDice loss and the regularization loss, our training objective also includes the FRE loss, NCC loss, and the cycle-consistency loss. An ablation study is conducted to analyze how these loss terms affect the performance of our networks.

### 3 Experiments

The 624 ears are partitioned into 465 ears for training, 66 ears for validation, and 93 ears for testing. The partition is random, with the constraint that ears of the same object cannot be used in both training and testing. We apply augmentation to the training set by rotating each image by 6 random angles in the range of -25 and 25 degrees about the x-, y-, and z-axis. The training images are blurred by applying a Gaussian filter with a kernel size selected randomly from {0, 0.5, 1.0, 1.5} with equal probability. This results in a training set expanded to 8835 images. Each image is clipped between its 5th and 95th intensity percentiles, and the intensity values are rescaled to -1 to 1. We use a batch size of 1, at each training step, 30% of the vertices on the ICA meshes are randomly sampled and used as the fiducial points for calculating the FRE loss.

## 4     Results

Figure 4 shows two cases for which our method leads to (a) good and (b) poor results. For each case, the first row shows three orthogonal views of the original atlas image in the atlas space. The second row shows the Post-CT image. The third row shows the atlas image registered to the Post-CT image. The fourth row shows the paired Pre-CT image of the Post-CT image. The warped atlas image (third row) should be as similar as possible to the Pre-CT image (fourth row). The last row shows the original segmentation mesh in the atlas image ($Mesh_{atlasSpc}$), the segmentation mesh in the Post-CT image generated using our method ($Mesh_{atlasSpc\text{-}postSpc}$), and the ground truth mesh in the Post-CT image ($Mesh_{postSpc}$). For $Mesh_{atlasSpc}$ and $Mesh_{postSpc}$, the ST, the SV, and the MD are shown in red, blue, and green, respectively. $Mesh_{atlas\text{-}post}$ is color-coded with the P2PE at each vertex on the mesh surfaces. Both these cases illustrate the severity of the artifact introduced by the implant. In the second case, the cochlea is barely visible.

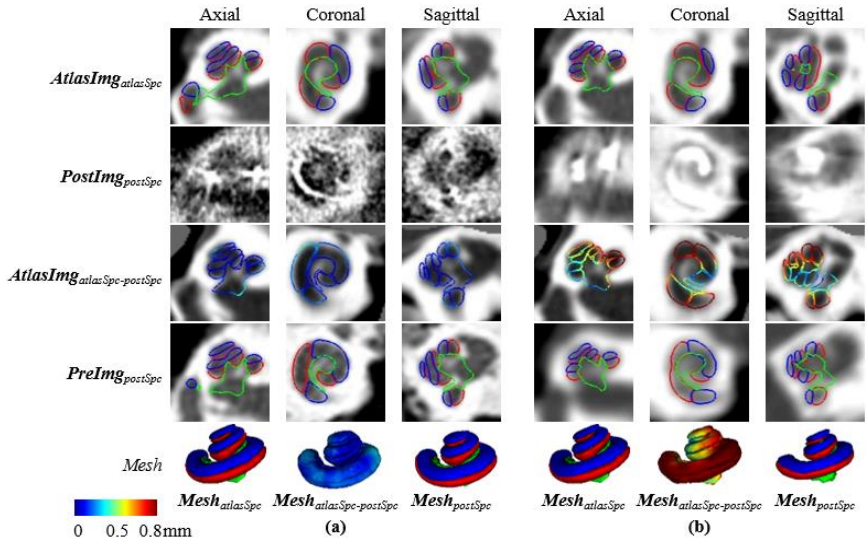

**Fig. 4.** Two example cases in which our method leads to (a) good and (b) poor results.

For each testing ear, we calculate the P2PEs of the vertices on the mesh surfaces of the ST, the SV, and the MD, respectively. We calculate the maximum (Max), median, and standard deviation (STD) of the P2PEs. Figure 5 shows the boxplots of these statistics for the 93 testing ears. "cGAN+ASM" denotes the results of the SOTA. "Proposed" denotes the results of our method. "Proposed-NoNCC", "Proposed-NoCycConsis", and "Proposed-NoFRE" denote the results of our proposed networks trained without using the NCC loss, the cycle-consistency loss, and the FRE loss. "Baseline" denotes the results of the baseline method. "No registration" denotes the P2PEs between the vertices on the mesh surfaces in the original atlas space and the Post-CT space. We perform two-sided and one-sided Wilcoxon signed-rank tests between the "Proposed" group and the other groups. The p-values have been corrected using the Holm-Bonferroni method [16]. The median values for each group are shown on top of the boxplots, in which red denotes that both the two-sided and the one-sided tests are significant, cyan denotes that only the two-sided test is significant, and blue denotes that the two-sided test is not significant. The results show that our networks trained using all of the proposed loss terms achieve a significantly lower segmentation error compared to the baseline method and the networks that are not trained using all of the loss terms. Our method produces results that are similar to those obtained with the SOTA in terms of the medians of the segmentation error. The Max of the segmentation error and the STD of the segmentation error for the SV and MD remain slightly superior to those obtained with the SOTA.

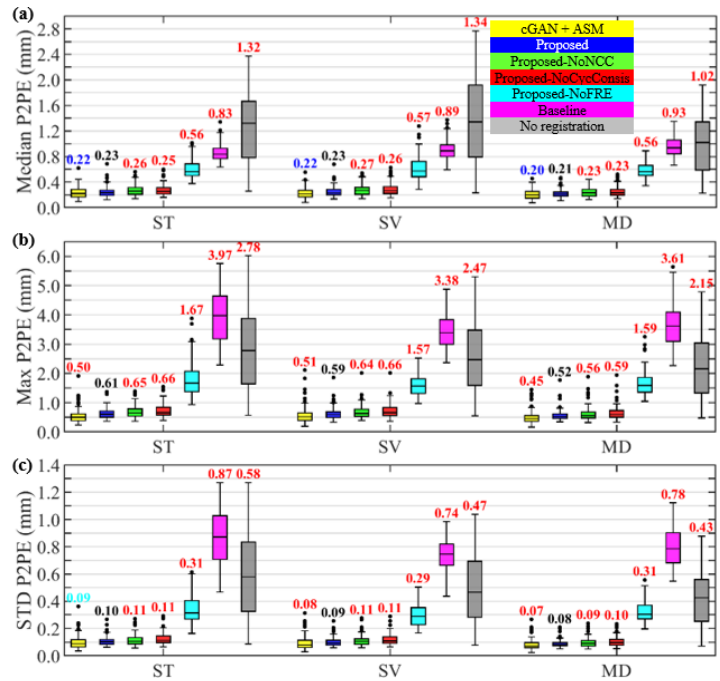

**Fig. 5.** Boxplots of (a) the median, (b) the Max, and (c) the STD of the P2PEs. A description of the numerical value color legend can be found in the text.

As mentioned earlier, the SOTA is a two-step process: (1) generate a synthetic Pre-CT image from a Post-CT image with cGANs trained for this purpose and (2) apply an ASM method to the synthetic image. Step 2 requires the very accurate registration of an atlas to the image to be segmented to initialize the ASM. This is achieved through an affine and then a non-rigid intensity-based registration in a volume-of-interest that includes the inner ear. Step 1 takes about 0.3s while step 2 takes on average 75s. The proposed method only requires providing a volume-of-interest that includes the inner ear to the networks and inference time is also about 0.3s. Segmentation is thus essentially instantaneous with the proposed method while it takes over a minute with the SOTA. This is of importance for clinical deployment and end-user acceptance.

## 5    Summary

We have developed networks capable of performing image registration between artifact-affected CT images and an artifact-free atlas image, which is a very challenging task because of the severity of the artifact introduced by the implant. Because we need to maintain point-to-point correspondence between meshes in the atlas and meshes in the segmented Post-CT images, we have introduced a point-to-point loss, which, to the best of our knowledge, has not yet been proposed. Our experiments have shown that this loss is critical to achieve results that are comparable to those obtained with the SOTA that relies on an ASM fitted to a preoperative image synthesized from a postoperative image. By design, ASM methods always produce plausible shapes. We have observed that with the point-to-point loss, our network also produces plausible shapes even when the images are of very poor quality (see Figure 4b). We hypothesize that, thanks to the point-to-point loss, the network has been able to learn the shape of the cochlea and can fit this shape to partial information in the post-operative image. More experiments are ongoing to verify this hypothesis.

## Acknowledgments

This work has been supported by NIH grants R01DC014037 and R01DC014462 and by the Advanced Computing Center for Research and Education (ACCRE) of Vanderbilt University. The content is solely the responsibility of the authors and does not necessarily represent the official views of these institutes.